# Plant Defense Multigene Families: I. Divergence of *Fusarium solani* -Induced Expression in *Pisum* and *Lathyrus*.

Sandhya Tewari[1,3], Stuart M. Brown[1,2] and Brian Fristensky[1]

[1]Department of Plant Science, University of Manitoba, Winnipeg, R3T 2N2, CANADA.

Current address: [2]Cell Biology Department, NYU Medical Center, 550 First Avenue, New York, NY 10016, U.S.A.

[3]Confederation of Indian Industry, 23 Institutional Area, Lodi Road, New Delhi - 110 003 INDIA

Corresponding author:

Dr. Brian Fristensky, frist@cc.umanitoba.ca

The defense response in plants challenged with pathogens is characterized by the activation of a diverse set of genes. Many of the same genes are induced in the defense responses of a wide range of plant species. How plant defense gene families evolve may therefore provide an important clue to our understanding of how disease resistance evolves. Because studies usually focus on a single host species, little data are available regarding changes in defense gene expression patterns as species diverge. The expression of defense-induced genes PR10, chitinase and chalcone synthase was assayed in four pea species (*Pisum sativum*, *P. humile*, *P. elatius* and *P. fulvum*) and two *Lathyrus* species (*L. sativus* and *L. tingitanus*) which exhibited a range of infection phenotypes with *Fusarium solani* . In *P. sativum*, resistance was accompanied by a strong induction of defense genes at 8 hr. post-inoculation. Weaker induction was seen in susceptible interactions in wild species. Divergence in the timing of PR10 expression was most striking between *P. sativum* and its closest realtive, *P. humile*. Two members of this multigene family, designated PR10.1 and PR10.2, are strongly-expressed in response to *Fusarium*, while the PR10.3 gene is more weakly expressed, among *Pisum* species. The rapidity with which PR10 expression evolves raises the question, is divergence of defense gene expression a part of the phenotypic diversity underlying plant/pathogen coevolution?

## INTRODUCTION

Molecular and genetic evidence support a two-tiered mechanism of induced plant defense in which resistance genes carry out signal transduction leading to the activation of defense genes [Dangl et al., 1995]. While many studies have examined the expression of genes associated with the defense response of plants to pathogens, these studies typically focus on a single host species or ecotype, or on differential lines isogenic for a single resistance locus. Little is known about whether patterns of defense gene expression are conserved among closely-related species. In general, protein coding sequences tend to be more highly conserved than non-coding sequences such as intron or promoter regions. Yet, if regulatory regions have more freedom to diverge, then their expression patterns might evolve rapidly as well. That is, even among closely-related species, or among ecotypes of a given species, the developmental or environmental contexts in which a gene is expressed could be quite varied.

Genes associated with inducible defense responses include those encoding enzymes of the phenylpropanoid pathway which are involved both in lignin production and synthesis of antimicrobial phytoalexins [Dixon and Paiva, 1995], as well as a growing list of "pathogenesis related (PR) proteins" [van Loon and van Kammen, 1970]. While the functions of many of the PR-proteins remain unknown [van Loon et al., 1994], others encode hydrolytic enzymes such as chitinases and β-1,3 glucanases [Bol et al., 1990; Boller et al., 1987; Bowels, 1990]. Considering the large number of defense genes, along with the fact that most of these genes are present as multigene families, [Harrison et al., 1995; Cramer et al., 1989; Koes et al., 1989; van Tunen et al., 1988; Corbin et al., 1987; Douglas et al., 1987], the divergence of expression patterns for these genes could affect host/pathogen compatibility. However, before this question can even begin to be addressed, it is first necessary to assess the degree to which defense gene expression is conserved among closely-related species. **If evolution of defense gene expression is part of host/pathogen coevolution, then it should be possible to detect changes in gene expression on at least as short a time scale as is required to detect changes in basic compatibility. Alternatively, if expression of a defense gene is strongly conserved among closely-related species, then the evolution of expression for that gene is unlikely to play a role in the evolution of basic compatibility among those species.**

*Pisum* and *Lathyrus* are members of the family Leguminosae, tribe Fabeae (=Vicieae) within the order Fabales [Waines, 1975]. *Pisum* consists of the garden pea *P. sativum* and three wild species, *P. humile, P.*

*elatius* and *P. fulvum* [Palmer et al., 1985]. *Pisum* species can be distinguished on the basis of morphologic, cytogenetic and molecular genetic data [Marx, 1977]. While *P. sativum*, *P. humile* and *P. elatius* have been known to form spontaneous hybrids [Ben-Ze'ev and Zohary, 1973], crosses between *P. fulvum* and other *Pisum* species result in seed set only when *P. fulvum* is the male parent. Additional data from electrophoretic patterns of albumin and globulin [Waines, 1975] and chloroplast DNA polymorphism in *Pisum* [Palmer et al., 1985] have led taxonomists to consider *P. fulvum* to be a distinct species and *P. sativum* to be an aggregate of *P. humile*, *P. elatius* and *P. sativum*. Within this aggregate, *P. humile* is considered to be the closest wild relative and the direct progenitor of cultivated pea.

Pod endocarp tissue as well as seedling tissue from *P. sativum* is susceptible to infection with the pea pathogen *Fusarium solani* f. sp. *pisi*. However, both tissues express a basic (non-host) resistance to the bean pathogen *F. solani* f. sp. *phaseoli*, in which germination and hyphal growth are inhibited [Christenson and Hadwiger,1973]. In addition to differences in pathogen growth, host responses such as an increase in phenylalanine ammonia lyase (PAL) activity and *de novo* synthesis of the phytoalexin pisatin, changes in host chromatin, and RNA synthesis [Hadwiger and Adams, 1978] are not only more rapid but also greater in intensity in response to the incompatible *F. solani* f. sp. *phaseoli* [Teasdale et al., 1974]. A marked increase in the rate of protein synthesis is also observed in endocarp tissue inoculated with *F. solani* f. sp. *phaseoli* , whereas *F. solani* f. sp. *pisi*-treated tissue shows only a slight increase [Christenson and Hadwiger, 1973]. Treatment with RNA and protein synthesis inhibitors within five hours post inoculation suppresses resistance to *F. solani* f. sp. *phaseoli*, whereas later treatments have no effect on resistance [Hadwiger, 1975; Teasdale et al., 1974].

Although the endocarp inoculation system offers conditions which are not typical of those existing in the field, the infection phenotypes with compatible and incompatible races of *F. solani* have been observed to be unaltered in pod tissues [Hadwiger et al., 1970]. In this assay, germination of the bean pathogen *F.solani* f. sp. *phaseoli* is inhibited while the pea pathogen *F. solani* f. sp. *pisi* germinates and grows [Teasdale et al., 1974]. Importantly, pod endocarp tissue serves as a large, uniform surface for inoculation on which all the cells are uniformly challenged.

Resistance of pea pod tissue to *F. solani* was previously demonstrated to be characterized by a suppression of germination or hyphal growth in the first few hours postinoculation. When pea pods are inoculated with the incompatible *F. solani* f. sp. *phaseoli*, macroconidiospores fail to germinate, and a yellow-green flourescence and a browning of the infection site indicative of a hypersensitive response is seen within 24 h.p.i. [Teasdale et al., 1974]. Resistance is preceeded by the increased accumulation of at least 21 "defense" proteins within 8 h.p.i [Wagoner et al., 1982]. When pods were heat-shocked at 40°C for 2hr. prior to a 6 hr. inoculation, extensive growth of the **incompatible** *F. solani* f. sp. *phaseoli* was seen [Hadwiger and Wagoner, 1983], defense proteins were suppressed and no hypersensitive response was evident by 24 h.p.i. When heat shock was followed by a 9 hr. recovery period, inhibition of fungal growth and expression of defense proteins were restored, although only a partial recovery of the hypersensitive response was seen. Interestingly, pod tissue heat-shocked after 6 h.p.i could still inhibit germination, although no hypersensitive response was evident at 24 h.p.i. These data suggest that suppression of spore germination requires an active response that occurs in the first 6 hr. after inoculation, and does not require hypersensitivity.

Differential screening of a cDNA library [Riggleman et al., 1985] prepared from endocarp tissue treated with *F. solani* f. sp. *phaseoli* was used to isolate "disease resistance response (Drr) cDNAs" [Fristensky et al., 1985]. Members of the Drr49 multigene family encode a 17 kD intracellular protein whose mRNA is induced by the elicitor chitosan, as well as *F. solani* f. sp. *phaseoli*. According to the nomenclature of van Loon *et al.*, [van Loon et al., 1994] this multigene family will henceforth be referred to as PR10. PR10 homologues have subsequently been identified as PcPR1 in parsley [Somssich et al., 1988], pathogenesis-related STH-2 in potato [Matton and Brisson, 1989], PvPR1 and PvPR2 in bean [Walter et al., 1990], AoPR1 in asparagus [Warner et al., 1993], and alfalfa [Esnault et al., 1993]; stress-induced SAM22 and H4 in soybean [Crowell et al., 1992]; the major birch pollen allergen *Betv*I [Breitender et al., 1989] and abscisic acid (ABA)-responsive ABR17 and ABR18 in pea [Iturriaga et al., 1994]. While the function of PR10 is not yet known, a protein isolated from *Ginseng* with 60-70% sequence identity with parsley PR10 was reported to have ribonuclease activity [Moiseyev et al., 1994].

The evolution of gene expression has seldom been specifically addressed in any experimental context, particularly not in plant/pathogen interactions. Therefore, all we attempt to accomplish in this study is to determine whether divergence in gene expression accompanies divergence in infection phenotype, between closely-related species. This will shed some light on the time scale needed for significant divergence in gene expression to occur. PR10 expression was assayed and compared to that of chitinase and chalcone synthase (CHS) in four pea species (*P. sativum*, *P. humile*, *P. elatius* and *P. fulvum*) and two *Lathyrus* species (*L. sativus* and *L. tingitanus*) which exhibited a range of infection phenotypes with *F. solani*. We show that resistance in *P. sativum* was accompanied by a strong induction of PR10 genes at 8 hr. post-inoculation, while susceptibility in wild legumes was associated with later or weaker induction. The PR10.1, PR10.2 subfamily was strongly-expressed in response to *Fusarium*, while the PR10.3 gene was much more weakly expressed, among *Pisum* species.

## MATERIALS AND METHODS

### Plant material and fungal strains

Wild accessions of *Pisum* (*P. humile* 713, *P. elatius* 721 and *P. fulvum* 706) used in this study were obtained from N. O. Polans, Northern Illinois University, U.S.A. *Lathyrus sativus* L720060 and *L. tingitanus* Nc 8f-3 were kindly provided by C. Campbell, Agriculture Canada Research Station, Morden, Canada. *P. sativum* c.v. Alaska was purchased from W. Atlee Burpee and Co., Warminister, PA. Strains of *Fusarium solani* f. sp. *pisi* and *F. solani* f.sp. *phaseoli* were obtained from American Type Culture Collection (Accession numbers 38136 and 38135 respectively). Cultures were grown and maintained on Potato Dextrose Agar (PDA) plates supplemented with a few milligrams of finely chopped pea leaf and stem tissue.

All the *Pisum* and *Lathyrus* plants were grown in growth rooms in pots in 2:1:1 soil:sand:peat mix under a day/ night cycle of 16/8 hours with temperatures of 22 /15 °C, respectively. The average light intensity using 1/3 0-lux wide spectrum to 2/3 cool white was 340 µ e m$^{-2}$ sec$^{-1}$.

### Pod inoculation procedure

Immature pods (less than 2 cm in length; five pods per treatment) having no developed seed were harvested from plants, slit longitudinally along the suture lines, and placed with the freshly opened side up in a sterile petri-dish. Fifty µl of a $10^6$ macroconidia/ml suspension was spread evenly on the pod. The plates were then incubated at room temperature under continuous florescent light and samples of the pod halves harvested at 8 and 48 hours. Pods treated with sterile distilled water served as controls.

### Staining and light microscopy

Inoculated pods were stained with 0.1% cotton blue (or trypan blue) in lactophenol (Anhydrous lactophenol 67% v/v; cotton blue 0.1 g w/v) for 30 sec., followed by a dip in distilled water. Pods were blotted dry on Kimwipes. Thin sections of endocarp tissue were prepared by slicing or sawing inoculated pods at a low angle, relative to the pod surface, using a scalpel with a #10 blade. Sections were wet-mounted with coverslips and photographed using Kodak Gold 100 film (GA135) on a photomicroscope (Carl Zeiss model # 63953). Pods were scored for resistance at 8 h.p.i. according to the criteria in Table 1. Five pods per treatment were examined. At least five fields on each pod were examined for scoring. Results from six independent experiments were averaged.

**Table 1.** Extent of hyphal proliferation on different host species.

|  | *F. solani* f. sp. *phaseoli* | | *F. solani* f. sp. *pisi* | |
| --- | --- | --- | --- | --- |
|  | 8 h.p.i. | 48 h.p.i. | 8 h.p.i. | 48 h.p.i. |
| *P. sativum* | - | - | + | # |
| *P. humile* | + | # | + | # |
| *P. elatius* | ++ | # | ++ | # |
| *P. fulvum* | +++ | # | +++ | # |
| *L. sativus* | ++++ | # | ++++ | # |
| *L. tingitanus* | +++++ | # | +++++ | # |

| Score | Light microscopy (8 hpi) | Appearance of pods (48 hpi) |
| --- | --- | --- |
| - | Less than 10% spores germinating; Germination tube less than 1/4th the size of the spore. | Light brown lesions; no maceration. |
| + | More than 50% spores germinating; Germination tube between 1/4 to 1/2 X the length of the spore. | Pinhead size dark brown lesions; little or no maceration of tissue. |
| ++ | More than 50% spores germinating; Germination tube ~1/2-1 X the length of the spore. | Pinhead size dark brown lesions; little or no maceration of tissue. |
| +++ | More than 50% spores germinating; Germination tube ~1-2 X the length of the spore. | Larger than pinhead size dark brown lesions; little or no maceration of tissue. |
| ++++ | More than 50% spores germinating; Germination tube ~2-3 X the length of the spore. | Large coalescing lesions; tissue macerated. |
| +++++ | More than 50% spores germinating; Germination tube more than 3 X the length of the spore. | Large coalescing lesions; tissue macerated. |
| # | Hyphal growth too dense to score. | Same as above. |

### DNA extraction and Southern blotting

Pea hypocotyls and young leaves were frozen in liquid nitrogen and ground to a fine powder using a mortar and pestle. One ml of extraction buffer [100 mM Tris-HCl (pH 8.0), 50 mM EDTA, 500 mM NaCl, 1.25% SDS] was added per 100 mg of tissue and incubated at 65 °C for 20 min. KOAc was added to a final concentration of 3 M and the samples were kept on ice for 20 min., followed by centrifugation at 12,000 g for 15 min. The supernatant was extracted twice with an equal volume of TE (10 mM Tris Cl pH 8.0, 1 mM EDTA)-equilibrated phenol. DNA was precipitated with 1 volume of isopropanol, reprecipitated with 2.5 vol. ethanol and 0.1 vol. 3 M NaOAc (pH 5.2), and the pellet dried and resuspended in TE.

For Southern blotting, 15 µg of genomic DNA from each species was digested with *Eco*RI, electrophoresed through 0.8% agarose in 1X 0.04M Tris acetate, 0.002 M EDTA (TAE) buffer, blotted onto Zeta probe GT membrane and UV crosslinked using the auto-crosslink mode of UV Stratalinker 1800 from Stratagene (1200 microjoules for 30 seconds). The blot was probed with PR10 probe (see "Preparation of probes"), except that 15 pg of pUC18 plasmid was included in the labelling reaction to detect the /*Hin*d III, pUC18/*Hin*f I marker.

### RNA extraction and Northern blotting

RNA was extracted from pods treated with fungus or water-treated controls at 8 and 48 h.p.i. RNA was extracted using a combination of the small-scale procedure for rapid isolation of plant RNAs [Verwoerd et al, 1989] and the phenol-chloroform method for RNA extraction [Ausubel et al., 1994]. Briefly, tissue was ground to a fine powder in liquid nitrogen, then mixed with hot (80 °C) extraction buffer [(1:1) phenol: ( 0.1 M LiCl, 100 mM Tris-HCl pH 8, 10 mM EDTA, 1% SDS)] to make a loose slurry (2-3 ml per g of tissue). One-half volume of chloroform was added and the suspension was mixed by vortexing. After centrifugation for 15 minutes at 975 g, the aqueous phase was removed to a fresh tube. One third volume of an 8 M solution of LiCl was added, then RNA collected by centrifugation for 10 min after overnight incubation at 4 °C. The RNA pellet was dissolved in 250 µl of Diethyl pyrocarbonate (DEPC) treated, sterile distilled water, reprecipitated with 0.1 volume of 3 M NaOAc pH 5.2 and 2.5 vol. of ethanol on ice for 20 min, centrifuged 20 min at 13,000 rpm (15,000 g), and the pellet redissolved in DEPC-treated sterile distilled water.

Ten micrograms total RNA was denatured using formaldehyde denaturation protocol [Ausubel et al., 1994] for RNA gel blot analysis and separated on 1.2% agarose-formaldehyde gels, blotted onto nylon membrane (Zeta-probe) using conditions recommended by the manufacturer and hybridized with $^{32}$P labelled, random primed probe in 0.25 M $Na_2HPO_4$, pH 7.2 and 7% SDS at 65°C. Filters were washed twice with 20 mM $Na_2HPO_4$, pH 7.2 and 5% SDS at 65 °C for 20 min.

### Recombinant plasmids

pI49KS and pI176KS consist of the pI49 (PR10.PS.1, GB::X13383) and pI176 (PR10.PS.2; GB::M18249) cDNAs, respectively [Fristensky et al., 1988], recloned between the *Sal* I and *Hin*d III sites of Bluescript KSm13+. pCC2 contains the PR10.PS.3 gene on a 3 kb *Sal* I fragment [Chiang and Hadwiger, 1990] subcloned into pUC18. p49cKS contains the 868bp *Nsi*I/*Xba*I coding sequence fragment from pCC2, recloned into *Pst*I/*Xba*I-digested BluescriptKSm13+. DC-CHIT-26 is a pea basic chitinase gene [Chang et al., 1995; GB:L37876] cloned between the CaMV 35S promoter and the NOS terminator in pBI121 (Clontech). pCHS2KS is the 1.6 kb pea chalcone synthase *Eco*R I fragment from pCHS2 [Harker et al., 1990] recloned into Bluescript KSm13+.

### Preparation of probes

All the probes were labeled with $\alpha^{32}$P-dCTP using the random primed DNA labelling system from GIBCO-BRL. Conserved PR10 probe was prepared from a PCR fragment amplified from pCC2 using conserved primers (oC49+3:cttactccaaaggttatt and oC49-5:taaggaacttctccttttac) which amplify all known PR10 genes in pea. The amplified band was isolated from agarose gel using Prep-A-Gene DNA purification matrix from Bio-Rad (Hercules, U.S.A.)

Chitinase probe was prepared by digesting DC-CHIT-26 with *Hin*d III and *Eco* RI to release the chitinase coding sequence along with CaMV 35S promoter and NOS terminator. The insert was gel-purified using Prep-A-Gene DNA purification matrix.

Chalcone synthase probe was made by labelling total pCHS2KS circular plasmid.

### Preparation of subfamily-specific probes

Probes specific for individual PR10 genes were generated by making use of a conserved *Bam*H1 restriction site near the 3' end of the protein coding region (140 bp 5' from the translational stop codon) of both the PR10.1 and PR10.3 genes. A second *Bam*HI site was present in the polylinker at the 3' end of the insert in PR10.1 plasmid allowing the isolation of roughly a 1 kb fragment containing the 3' coding sequence and 3' flanking DNA. In the PR10.3 clone (pCC2), a second *Bam*HI site was present in the insert at 480 bp 3' of the stop codon, allowing the isolation of a 716 bp *Bam*HI fragment containing the 3' end of the coding sequence and 3' flanking DNA. Both fragments were separated by gel electrophoresis, cut from the gel and recovered from the gel slice using the Prep-A-Gene kit from Bio-Rad. The recovered fragments were labelled according to the method described above.

### Preparation of Markers

Marker PR10.1 was a mixture of equimolar amounts of pI49KS (PR10.1) digests with *Pst* I (3343, 426), *Hin*d III (3769) and *Hin*d III/*Xho* I double digest (2943, 826). The numbers in parentheses represent the size in base pairs of fragments released. The underlined fragments represent the bands that hybridize with PR10 subfamily-specific probes.

Marker PR10.2 was prepared by mixing equimolar amounts of the following pI176KS (PR10.2) digests: *Pst* I (<u>3336</u>, 427), *Hin*d III (<u>3763</u>), *Pvu* II (<u>2519</u>, <u>1244</u>) and *Hin*d III/*Xho* I double digest (<u>2943</u>, 820).

Marker PR10.3 was prepared by mixing equimolar amounts of the following p49cKS digests: *Hin*dIII (<u>3806</u>), *Hin*dIII/*Pst*I (<u>3283</u>, <u>523</u>) and *Hin*d3/*Sac*I (2888, <u>918</u>).

Marker M was prepared by mixing separate digests of lambda DNA with *Hin*d III and pUC19 with *Hin*f I.

One nanogram of each DNA marker was denatured in formaldehyde as described above prior to loading on formaldehyde gels.

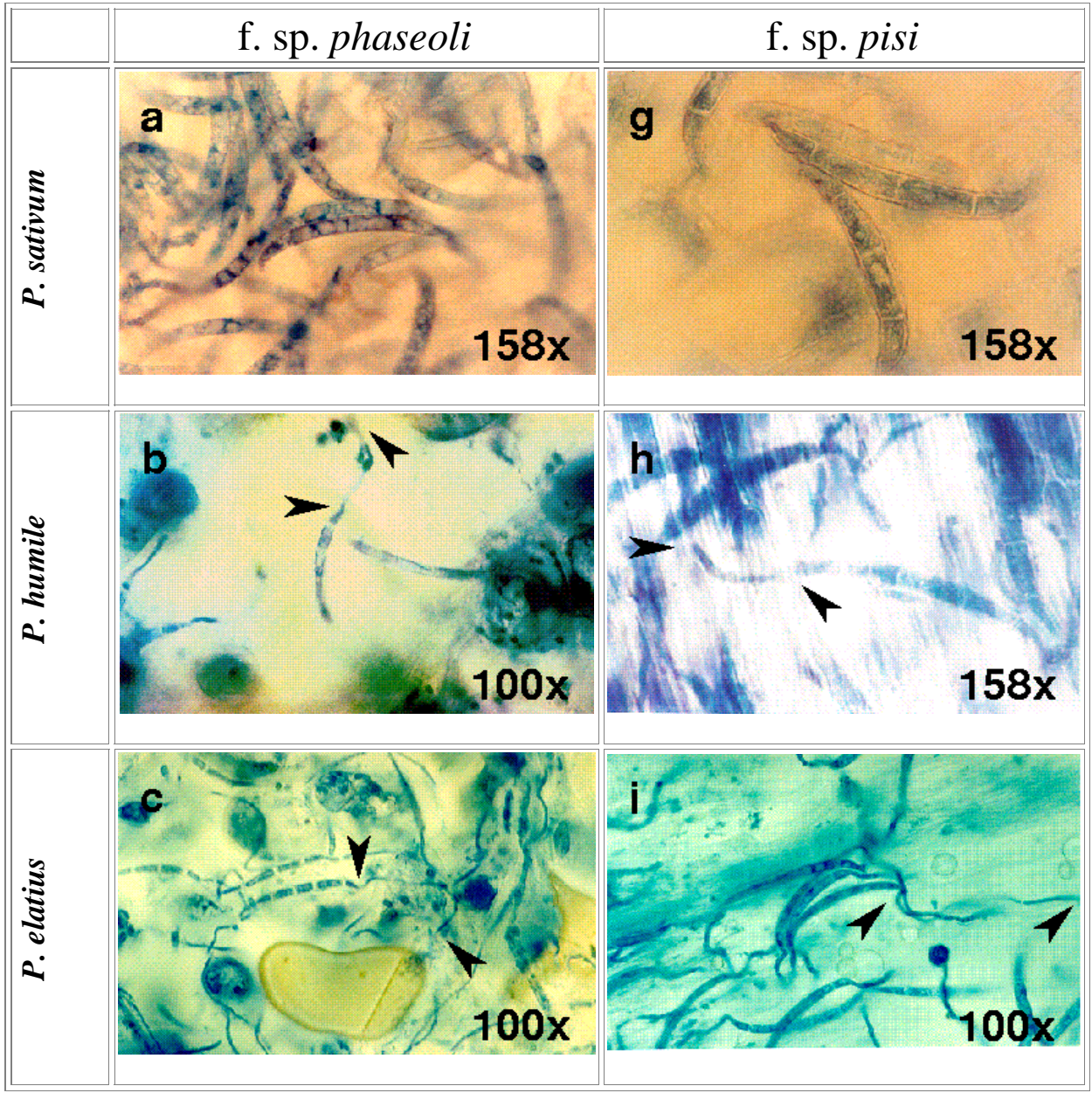

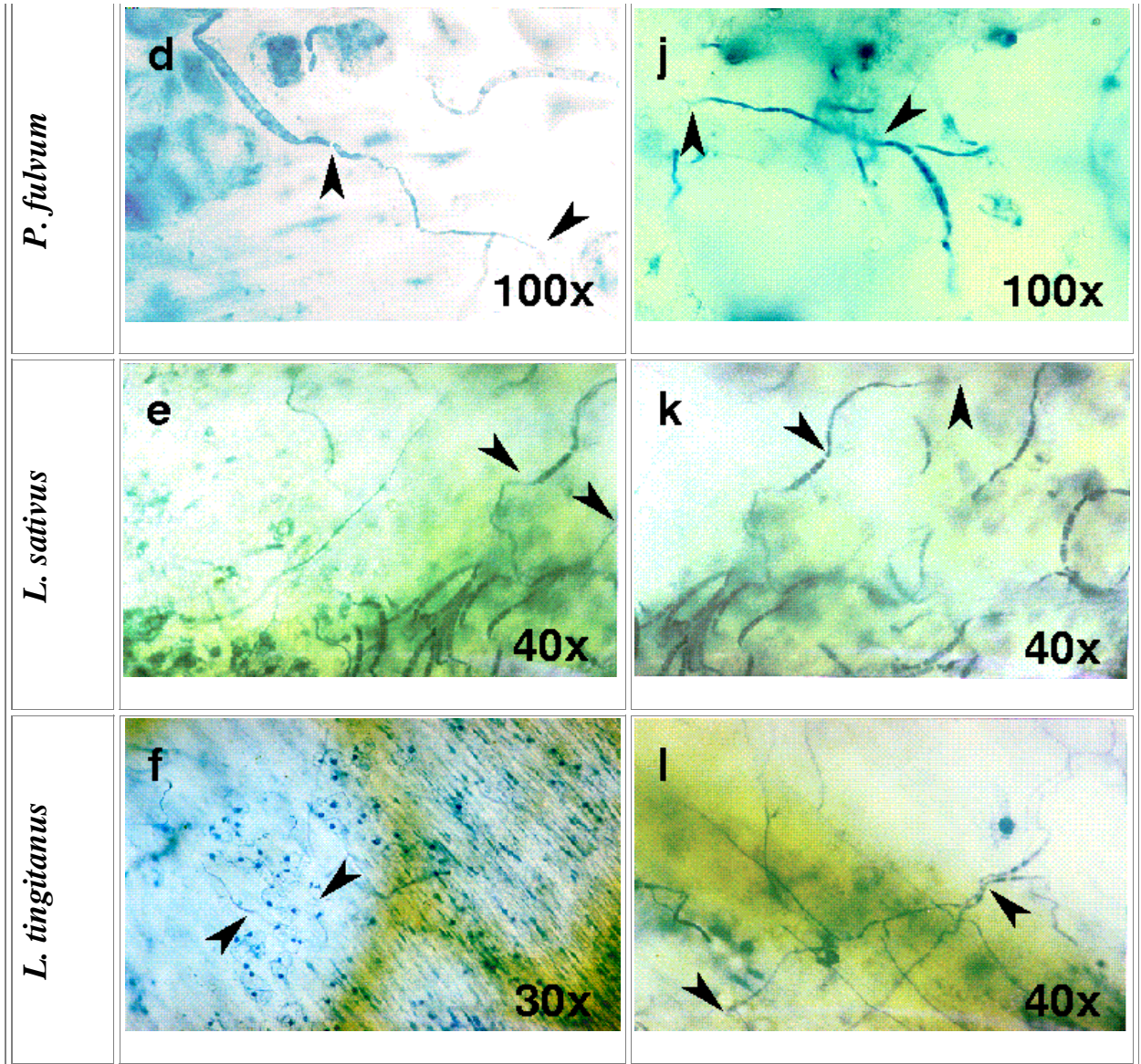

**Figure 1.** Light micrographs of *Fusarium solani* f. sp. *phaseoli* (**a-f**) and *F. solani* f. sp. *pisi* (**g-l**) macroconidia on the endocarp tissue of *Pisum* and *Lathyrus* species at 8 hours post inoculation. Panels a-f show f. sp. *phaseoli* on *P. sativum* (**a**), *P. humile* (**b**), *P. elatius* (**c**), *P. fulvum* (d), *L. sativus* e) and *L. tingitanus* (**f**). Panels **g-l** show f. sp. *pisi* on *P. sativum* (g), *P. humile* (h), *P. elatius* (**i**), *P. fulvum* (**j**), *L. sativus* (k) and *L. tingitanus* (l). Interactions are arranged in increasing infection phenotype score (Table 1) from top to bottom. Magnifications are indicated at the lower right corner of each plate. Arrows indicate the interval between the spore apex and the hyphal tip. Due to unevenness of the endocarp surface, entire hypha can seldom be visualized in a single focal plane.

## RESULTS

**Divergence of infection phenotypes in *Pisum* and *Lathyrus***

As described in the heat shock experiments cited in the Introduction, inhibition of macroconidiospore germination requires an active response by pea tissue within the first 6 hours after inoculation. Therefore, this work focuses on the early hours postinoculation. Compatibility of *F. solani* races with *Pisum* or *Lathyrus* species was measured with respect to percent spore germination and extent of hyphal growth at 8 h.p.i, as described in methods. Results are summarized in Table 1. Figure 1a shows that on *P. sativum*, the incompatible *F. solani* f. sp. *phaseoli* does not germinate, while f. sp. *pisi* exhibits germination but very little hyphal growth by 8 h.p.i (Fig. 1g).

Wild *Pisum* and *Lathyrus* permitted more hyphal proliferation than domestic pea (Table 1). The delay in hyphal growth at 8 h.p.i. was less pronounced in these species than in *P. sativum* (Fig. 1). The closest relative of garden pea, *P. humile*, inhibited both pathogens, albeit more weakly than *P. sativum*. Germ tubes at 8 h.p.i. were about ¼-½ the size of the spores (Fig.1 b and h). *P. elatius* and *P. fulvum* were even more permissive to hyphal growth of both the pathogens, with scores of ++ and +++ respectively (Table 1, Fig 1c, d, i and j). Both *Lathyrus* species allowed extensive growth of both pathogens with germ tubes more than twice the length of the spore (Fig. 1e, f, k and l) within this same period.

The ratings in Table 1 are averaged results from six experiments, representing the majority of spores scored for a given treatment. Notwithstanding, two observations must be made. First, in all treatments some ungerminated spores could be found, even in cases such as the interaction of *L. sativus* with *F. solani* f. sp. *phaseoli* in which the vast majority of spores had extensive hyphal growth by 8 h.p.i. Secondly, on all hosts except *P. sativum*, a small percentage of spores appeared to completely escape suppression of hyphal growth, with hyphae 3 or more times the length of the spore.

**PR10 is present as a multigene family in *Pisum* and *Lathyrus***

In order to confirm the presence of homologous PR10 sequences in *Pisum* and *Lathyrus* species, the *P. sativum* PR10.1 gene was used as a probe in a DNA gel blot of *Pisum* and *Lathyrus* species (Fig. 2). Band patterns in all species were consistent with 3-5 gene copies per haploid genome, demonstrating the existence of PR10 multigene families in each species. *P. sativum* and its closest relative, *P. humile*, share a common 9.4 kb band, while an 8.0 kb band is common to all *Pisum* species except *P. humile*. *P. humile* and *P. fulvum* share a 3.4 kb band. No bands appear to be conserved between *Pisum* and

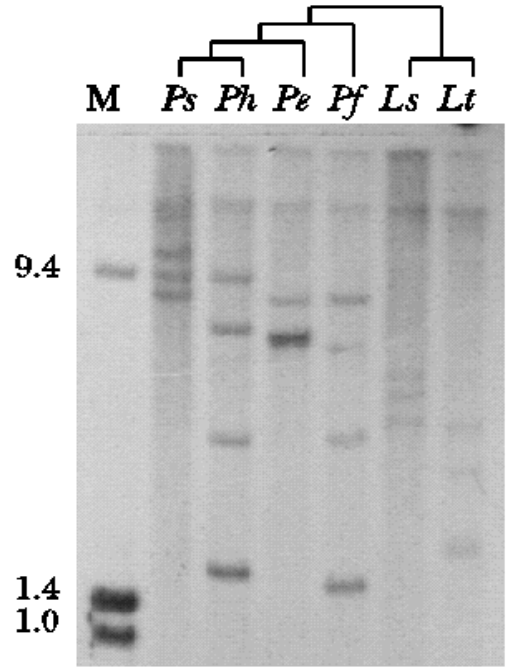

**Figure 2.** Genomic DNA gel blot analysis of *P. sativum* (**Ps**), *P. humile* (**Ph**), *P. elatius* (**Pe**), *P. fulvum* (**Pf**), *L. sativus* (**Ls**) and *L. tingitanus* (**Lt**) genomic DNA using $^{32}$P-labelled PR10.1 probe. The relationships between taxa, as described in the introduction, are represented in a cladogram. Fifteen micrograms of *Eco* RI-digested genomic DNA was loaded in each lane. **M** = Lambda/*Hin*d III, pUC19/*Hin*f I marker.

*Lathyrus*. Finally, the lower band intensity seen in the *Lathyrus* lanes suggests that *Pisum* and *Lathyrus* PR10 genes have diverged substantially. The conservation of bands within *Pisum*, but not between *Pisum* and *Lathyrus*, is consistent with the fact that between-species divergence has been more recent than the divergence of *Pisum* and *Lathyrus*. The interfertility between *Pisum* species, although partial [Waines, 1975], may also have contributed to the observed interspecific band conservation.

**Divergence of gene expression patterns**

*P. sativum*

In *P. sativum*, which is resistant to *F. solani* f. sp. *phaseoli*, PR10 mRNA was present at high levels within 8 h.p.i. (Fig. 3) but decreased in abundance by 48 h.p.i. A similar pattern was observed with CHS and chitinase genes but the signal was much weaker than that of PR10 (Fig. 3).

In contrast, *P. sativum* inhibits the germination of *F. solani* f. sp. *pisi* spores at 8 h.p.i. although by 48 h.p.i., the fungus is observed to grow uninhibited. At 8 h.p.i., PR10 was observed to be induced to a high level

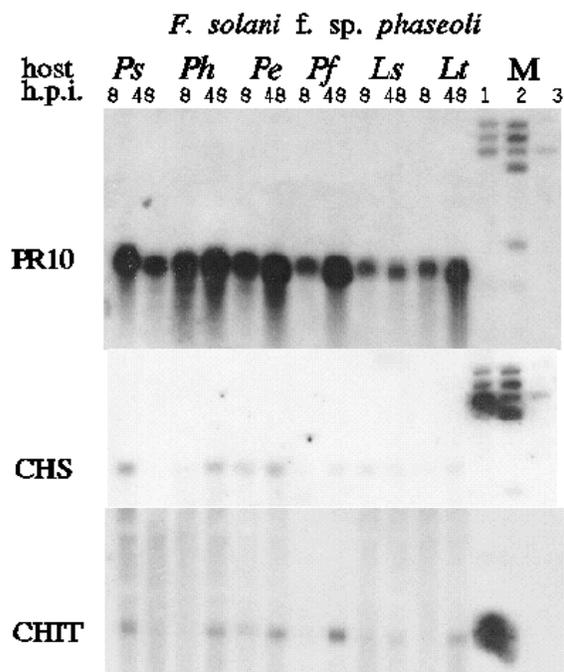
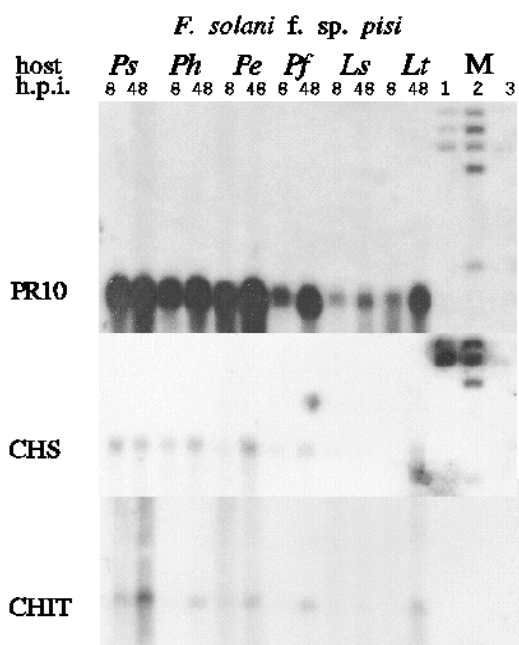

**Figure 3.** Expression of PR10, CHS and chitinase (CHIT) mRNA in pod tissue of *Pisum* and *Lathyrus* species inoculated with *F. solani* f. sp. *phaseoli* for 8 or 48 hours. RNA gel blots (5 μg per lane) were probed with $^{32}$P-labelled PR10, CHS and chitinase probes. The same filters were sequentially stripped and reprobed to maintain consistency between experiments. **Ps** = *P. sativum*, **Ph** = *P. humile*, **Pe** = *P. elatius*, **Pf** = *P. fulvum*, **Ls** = *L. sativus* and **Lt** = *L. tingitanus*. **M -1, -2, -3** = PR10.1, PR10.2 and PR10.3 markers, as described in the methods section. **h.p.i.** = hours post inoculation. Because the CHS probe was made from total plasmid, some marker bands hybridize, whereas the CHIT probe was made from isolated insert, resulting in no marker hybridization. The large spot at the right of the CHIT figure is an artifact.

**Figure 4.** Expression of PR10, CHS and chitinase mRNA in pod tissue of *Pisum* and *Lathyrus* species inoculated with *F. solani* f. sp. *pisi* for 8 and 48 hours. All other experimental conditions and annotations are the same as in Fig. 3.

*P. elatius*
*P. elatius* allowed moderate growth of both *F. solani* f. sp. *phaseoli* and f. sp. *pisi* (Table 1). In response to both pathogens, PR10 was expressed to high levels within 8 h.p.i. with the expression increasing by 48 h.p.i (Figs. 3 and 4). A similar pattern was observed for chitinase and CHS with both pathogens although transcript abundance was much lower (Figs. 3 and 4).

*P. fulvum*
Both *F. solani* f. sp. *phaseoli* and f. sp. *pisi* were able to grow relatively uninhibited on *P. fulvum* (Table 1). It showed a remarkably similar expression pattern for all three genes in response to both pathogens. This pattern was characterized by very low to undetectable expression at 8 h.p.i. followed by relatively higher transcript accumulation at 48 h.p.i. (Figs. 3 and 4).

*L. sativus*
*L. sativus* allowed both fungi to germinate and grow rapidly (Table 1). PR10 expression was somewhat greater at 48 h.p.i than 8 h.p.i, while CHS and chitinase transcripts were barely detected in response to either pathogen (Figs 3 and 4). This does not necessarily imply low expression of these genes in *L. sativus*. It is possible that the latter two pea probes in response to this pathogen (Fig. 4). However, unlike that with *F. solani* f. sp. *phaseoli*, expression of PR10 was maintained at high level up to 48 hour. CHS mRNA was much less abundant than PR10 but exhibited the same pattern at both time-points (Fig. 4). Chitinase mRNA was also detectable within 8 h.p.i. and its level rose by 48 h.p.i.

*P. humile*
*P. humile* which partially inhibited both pathogens (Table 1), also expressed PR10 to high levels at 8 h.p.i. in response to *F. solani* f. sp. *phaseoli*, albeit lower than that in *P. sativum* (Fig. 3). CHS and chitinase mRNA were barely detectable in *P. humile* at 8 h.p.i. but appeared by 48 h.p.i.

In response to *F. solani* f. sp. *pisi*, PR10 transcript was abundant at 8 h.p.i., accumulating to higher levels by 48 h.p.i. (Fig. 4). Chitinase and CHS mRNAs exhibited a stronger signal at 48 h.p.i than at 8 h.p.i.

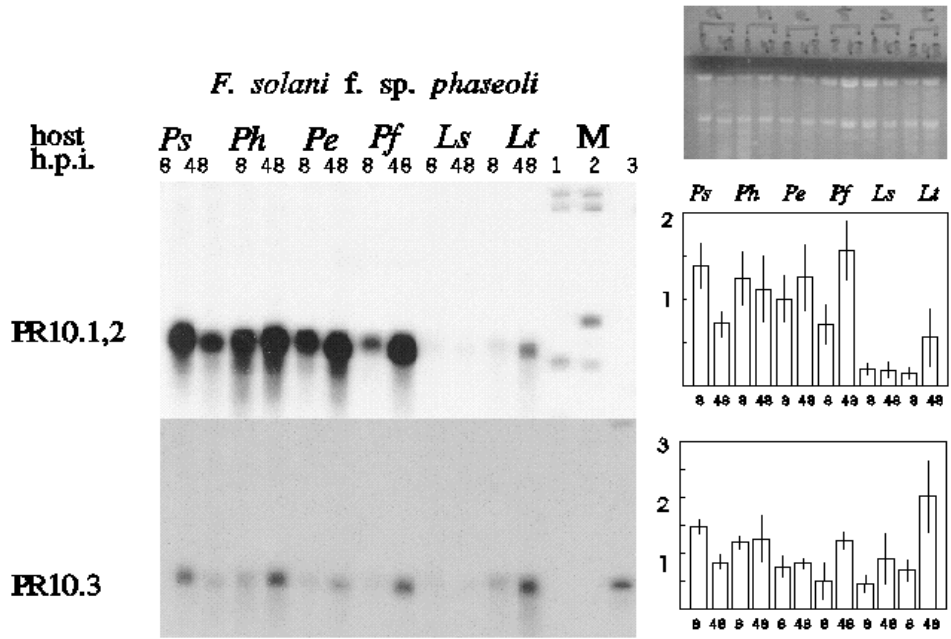

**Figure 5.** Differential expression of PR10 subfamilies in *Pisum* and *Lathyrus* species in reponse to inoculation with *F. solani* f. sp. *phaseoli* using the subfamily-specific probes derived from the 3' untranslated region of the genes as described in Methods. Total RNA is visualized in ethidium bromide staining in the upper right panel. All other experimental conditions and annotations are the same as in Fig. 3. Histograms to the right of gel images represent the means of normalized signal from autoradiograms as measured by densitometry. Each histogram represents the mean of at least 3 experiments. The standard error of the mean is indicated by vertical lines superimposed on each bar. Since autoradiographic signals for many values fell outside the linear response range of the film, the histograms underestimate the differences between treatments.

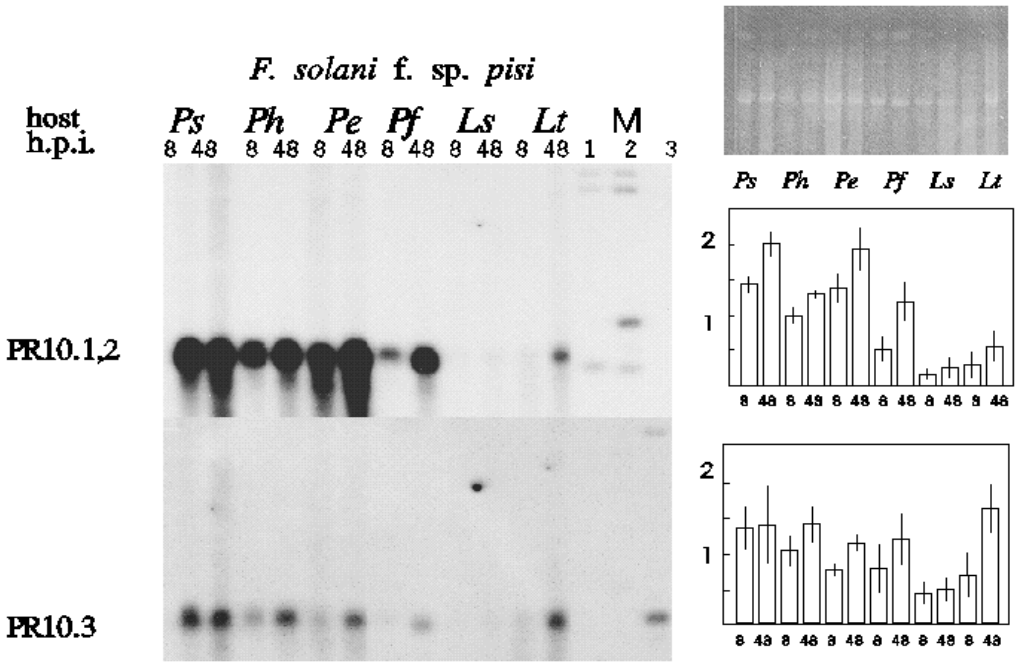

**Figure 6.** Differential expression of PR10 subfamilies in *Pisum* and *Lathyrus* species in reponse to inoculation with *F. solani* f. sp. *pisi* using the subfamily-specific probes derived from the 3' untranslated region of the genes as described in Methods. All other experimental conditions and annotations are the same as in Fig. 5. Note that the upper ribosomal bands in the EtBr-stained gel appear faint, due to quenching by xylene cyanol dye.

hybridize only weakly due to lack of sequence conservation between *Pisum* and *Lathyrus*.

*L. tingitanus*

In *L. tingitanus,* which allowed maximum fungal growth among all the tested host species (Table 1), PR10 transcript was detectable by 8 h.p.i (Figs 3 and 4), accumulating to higher levels by 48 h.p.i. CHS RNA was hardly detectable in this species (Figs 3 and 4). Chitinase was not detectable at 8 h.p.i with either pathogen but some transcript accumulation was observed at 48 h.p.i. in response to both pathogens.

**Differential expression of PR10 subfamilies**

Conservation of distinct PR10 subfamilies within *Pisum* and *Lathyrus* species prompted us to question if expression patterns for PR10 subfamily members are consistent throughout *Pisum,* or whether these patterns change along with the observed changes in germination and hyphal growth. Subfamily-specific probes were constructed from the C-terminal protein coding regions, extending into the 3' non-transcribed region of each gene (see Methods). These probes were then used in gel blots using RNA isolated from different host species inoculated with *F. solani* f. sp. *phaseoli* or f. sp. *pisi* to determine if each subfamily was active in different hosts. The specificity of these probes was verified by the use of plasmids containing PR10.1, PR10.2 and PR10.3 sequences as internal controls on each RNA blot. In Figures 5 and 6, the PR10.1 probe hybridized to PR10.1 and PR10.2, but not to PR10.3. The PR10.3 specific probe hybridized only to the PR10.3 plasmid. The stronger signal with the PR10.1 probe as compared to the PR10.3 probe indicates that PR10.1/PR10.2 subfamily specific transcripts accumulate in greater abundance, as compared to PR10.3 transcripts, in *Pisum* and *Lathyrus* inoculated with *F. solani*. Low signal in *Lathyrus* under higher stringency hybridization and washing conditions indicates that PR10 genes have diverged substantially and is consistent with low signal in the DNA gel blot (Fig 2).

Expression patterns seen with PR10 subfamily-specific probes (Fig. 5 & 6) generally agreed with results using non-specific PR10 probes (Fig. 3 & 4). **In** *P. sativum* pods inoculated with *F. solani* f. sp. *phaseoli* (Figs 3, 5) expression of PR10.1/PR10.2 and PR10.3 is stronger at 8 h.p.i than at 48 h.p.i. In *P. humile*, expression at 8 and 48 h.p.i. are fairly uniform, although in some experiments greater expression was seen at 8 h.p.i. In *P. elatius*, *P. fulvum* and *L. tingitanus* expression at 48 h.p.i. is stronger than at 8 h.p.i. In tissue inoculated with *F. solani* f. sp. *pisi* (Figs. 4, 6) expression of PR10 genes is typically weaker at 8 h.p.i. than at 48 h.p.i. One difference between results obtained with the non-specific PR10 probe, versus the subfamiliy-specific probes, is that with the PR10.1,2 probe, signal for *Lathyrus* species is much weaker than for *Pisum* species. Stronger signals were obtained in *Lathyrus* using non-specific probes. This result is not surprising, since the subfamily-specific probes contain only the C-terminal part of the coding region, as well as the 3' untranslated region, which are likely to be the most divergent, between species. This would be consistent with results in Fig. 2, in which weaker autoradiographic signal is also seen in hybridization with *Lathyrus* genomic DNA, as compared to *Pisum* DNA. Interestingly, signal intensities using the pea PR10.3 probe are comparable in both *Pisum* and *Lathyrus*.

## DISCUSSION

How plant defense gene families evolve may provide an important clue to our understanding of how disease resistance evolves. In order to study the evolution of defense gene expression, it was necessary to first determine whether infection phenotype differed within a set of closely-related species. Since it is not possible to directly observe speciation in progress, the best alternative is to study a range of species, some of which are partly interfertile, and others which are not. At the same time, few papers in the plant pathology literature examine interactions in wild plant species, or compare a resistance response in a domestic plant with that in a wild plant. This is an important point, because the strong bias towards domestic species is undoubtedly skewing our picture of host/pathogen interactions. For both of these reasons, we selected *Pisum* and *Lathyrus* species for this study.

Pod endocarp tissue from *P. sativum* inhibited the germination of macroconidia of *F. solani* f. sp. *phaseoli*. *P. humile*, which is most closely-related to *P. sativum*, exhibited a phenotype more similar to *P. sativum* than the other two wild species with a relatively strong inhibition to germination of *F. solani* f. sp. *phaseoli* spores. *Lathyrus* species, which are further diverged from *Pisum*, were more permissive to hyphal growth. A similar increase in compatibility was seen in the interaction with the pea pathogen, *F. solani* f. sp. *pisi*. While these experiments do not specifically examine variation of defense response within each species, it is worth noting that the divergence of interaction phenotype appears to be gradual because neighbouring species always had the most similar scores.

Changes in the interaction phenotype across species were accompanied by divergence of expression patterns for PR10, CHS and chitinase genes. Using the same basic chitinase probe, Chang *et al.* [1995] also detected induction of chitinase in response to *F. solani*, while CHS expression has not previously been studied in this pathosystem. While there are some similarities between the PR10 pattern of expression and those of CHS and chitinase, there are also some apparent differences in the timing and levels of respective transcript accumulation (Fig. 3 & 4). Thus, while some regulatory pathways may be shared among these gene families, our data do not point to a strict coordinate regulation.

Resistance was accompanied by expression of defense genes at 8 h.p.i. In *P. sativum, P. humile* and *P. elatius*, significant accumulation of PR10 occurs within 8 h.p.i. All three species show fewer than 50% spore germination for both pathogens, as well as little or no hyphal growth at this time (Fig. 1). In *P. fulvum*, *L. sativus* and *L. tingitanus*, which allow greater than 50% germination and extensive hyphal growth by 8 h.p.i., there is substantially less PR10 mRNA accumulation at 8 hours.

All species except *P. sativum* show a similar pattern of expression of PR10 genes during infection with either *F. solani* f. sp. *pisi* or *F. solani* f. sp. *phaseoli*. This pattern is characterized by either a weak or moderate signal in the first 8 h.p.i., followed by a stronger induction by 48 hours. In contrast, *P. sativum* shows a high accumulation of PR10 transcript at 8 hours after infection with either pathogen, followed by a decline in transcript levels by 48 hours in case of *F. solani* f. sp. *phaseoli*, but similar levels of expression at both time points after infection with *F. solani* f. sp. *pisi*. These results parallel the observation that on *P. sativum* tissue inoculated with *F. solani* f. sp. *phaseoli*, hyphal growth was completely suppressed, whereas on tissue inoculated with *F. solani* f. sp. *pisi*, growth is initiated, but is halted, to resume at later times.

Pea PR10 hybridized to multiple bands in the *Eco* RI digested genomic DNA from wild *Pisum* and *Lathyrus*, indicating that PR10 exists as a multigene family in these taxa. RNA gel blot analysis using PR10 subfamily-specific probes showed that PR10.1/2 subfamily transcripts increased greatly in response to *F. solani* while that of PR10.3 subfamily ranged from weak to undetectable in all species. Mylona *et al.*, [1994] have independently cloned the pea PR10.3 cDNA while isolating genes expressed in root epidermis and root-hairs. PR10.3 (referred to as RH2 in that paper) transcript was far more abundant in roots than transcripts detected using PR10.1-specific oligonucleotides. Further, inoculation of roots with *Rhizobium leguminosarum* bv. *viciae* did not have any detectable effect on the already high PR10.3 transcript accumulation, but caused a slight increase in accumulation of PR10.1 transcript over control levels. Savouré et al. [1997] demonstrated that PR10 genes in the legume *Medicago sativa* are induced by Nod (nodulation) factors in suspension culture, but expressed constitutively in roots. In contrast, Gamas et al. [1996] have identified PR10 genes in *Medicago truncatula* that are induced during nodule development, but not expressed in roots. While the latter two studies did not use gene-specific probes, they do provide further evidence that gene expression patterns for PR10 genes change from species to species, both with respect to development and to plant/microbe interactions.

To our knowledge, the divergence of defense gene regulation has not been compared among other groups of closely-related species. Therefore, it is not known whether the evolution of defense gene regulation in general is as dynamic as that seen in this study. It is commonly observed that non-coding sequences evolve more rapidly than protein coding sequences. For example, 3' untranslated regions of genes are often used as gene-specific probes due to their characteristic lack of conservation, relative to translated regions [Dean et al., 1985]. However, it is not known whether regulatory sequence divergence is responsible for the divergence in PR10 gene expression observed here. Another posibility is that changes in signal transduction pathways, perhaps even the same pathways leading to the observed changes infection phenotype, are responsible for divergence in PR10 gene expression patterns.

Since dozens of genes may be involved in the defense response, and most of these are present as multigene families, the precise set of genes activated in response to a given pathogen, and their patterns of regulation, could vary enormously, within and between species. As a consequence, the phenotypic diversity of plant populations, with respect to their response to pathogens, may be greater than revealed by typical gene expression studies.

Demonstrating a causal link between the changes in basic compatibility between plant and pathogen and the evolution of defense multigene families is beyond the scope of any single study such as this. In this paper, only three out of the multitude of known defense genes were studied. However, it is fair to say that one component underlying that phenotypic diversity may be differential regulation of genes within multigene families encoding defense proteins.

## ACKNOWLEDGEMENTS

We would like to thank Dr. Enrico Coen for clone pCHS2 clone, Dr. Lee Hadwiger clones pCC2 and DC-CHIT-26 and Dr. J. Davie for use of densitometry equipment. Also, thanks to Dr. Tom Warkentin and Dr. Glen Klassen for comments on the manuscript. This work was supported by Operating Grant OGP0105628 from the Natural Sciences and Engineering Research Council of Canada. Support was